# THE HEATED LAMINAR VERTICAL JET IN A LIQUID WITH POWER-LAW TEMPERATURE DEPENDENCE OF DENSITY

V. A. Sharifulin

Perm State Technical University, Perm, Russia

e-mail: sharifulin@perm.ru

## 1. Introduction

Water layer near $4°C$ is a peculiar example of convectional system with non-uniform stratification. Anomalous heat extension is well known to takes place in this area. At $T_i = 4°C$ water density is maximum, in the interval from melting point $T = 0°$ to $T = 4°C$, water behavior is anomalous: water density increases with the increase of temperature; for $T > T_i$ density decreases with the increase of temperature. In the temperature interval $0-8°C$, temperature dependence of density can be approximated with high degree of accuracy by symmetric parabola with its maximum at the point of inversion. In a wider range of parameters the water behavior is described by power-law temperature dependence of density as follows:

$$\rho(T) \sim 1 - \beta |T - T_i|^\gamma.$$

The problem of a heated vertical jet of an incompressible liquid in a boundary layer was first investigated analytically and numerically by Brand & Lahey [1]. Analytical solutions were obtained for the certain values of Prandtl number by similarity methods. Mollendorf at al. [2] numerically investigated the problem for the case of power-law temperature dependence of density corresponding to the water behavior for normal pressure and temperature near $4°C$. M.A Goldshtik, V. N. Shtern & N. I. Yavorsky [3] treated the jet problem for cubic temperature dependence of density. In their work the authors used standard Bussinesq equations without boundary-layer approximation. The free jet behavior, both numerically and analytically with small parameter expansion method, were investigated.

One can see that free jet problem attracts constant interest of investigators. Although from short review above follows necessity of analytical solution account for the case of liquids with power-law temperature dependence of density. And the present study serves to improve completeness of the problem in this way.

Analytical solutions for the certain values of Prandtl number and arbitrary values of inversion rate $\gamma$ were found by similarity methods same to that employed by Brand & Lahey [1].

## 2. The two-dimensional jet

A system of rectangular co-ordinates $(x, y)$ is chosen such that the $x$-axis coincides with the symmetry axis of the jet. The boundary-layer equations, expressing the conservation laws of mass, momentum, and energy are:

$$\frac{\partial v_x}{\partial x} + \frac{\partial v_y}{\partial y} = 0, \tag{2.1}$$

$$v_x \frac{\partial v_x}{\partial x} + v_y \frac{\partial v_x}{\partial y} = \nu \frac{\partial^2 v_z}{\partial y^2} + g\alpha T_\infty^\gamma \theta^\gamma, \tag{2.2}$$

$$\sigma\left(v_x \frac{\partial \theta}{\partial x} + v_y \frac{\partial \theta}{\partial y}\right) = v \frac{\partial^2 \theta}{\partial y^2}. \tag{2.3}$$

In these equations, $v_x$ and $v_y$ are velocity components, and $\theta$ is a dimensionless temperature difference related to the local temperature, $T(x, y)$ and to the temperature far from the jet, $T_\infty$ by

$$\theta = (T - T_\infty)/T_\infty \tag{2.4}$$

Other constants appearing in the equations are: $v$ the kinematic viscosity; $\beta$, the coefficient of volume expansion; $g$, the acceleration of gravity; and $\sigma$, the Prandtl number.

Boundary conditions to be satisfied are:

$$at \ y = 0, \ v_y = \frac{\partial v_x}{\partial y} = \frac{\partial \theta}{\partial y} = 0; \tag{2.5}$$

$$at \ y = \infty, \ v_x = \theta = 0 \tag{2.6}$$

The continuity equation (2.3) implies the existence of a stream function. $\psi(x, y)$, Such that

$$v_x = \partial \psi / \partial y, \quad v_y = -\partial \psi / \partial x \tag{2.7}$$

The partial differential equations are reduced to ordinary equations by means of the following transformation:

$$\eta = ayx^{-\frac{1+\gamma}{4+\gamma}}, \tag{2.8}$$

$$\psi(x, y) = avx^{\frac{3}{4+\gamma}} f(\eta) \tag{2.9}$$

$$\theta(x, y) = \left(a^4 v^2 \left(g\beta T_\infty^\gamma\right)^{-1} \tau(\eta)\right)^{1/\gamma} x^{-\frac{3}{4+\gamma}} \tag{2.10}$$

The arbitrary constant, $a$, is included so as to make $\eta, f(\eta)$ and $\tau(\eta)$ dimensionless. It can be chosen so as to match the mathematical solution to a particular physical case.

The governing equations for the problem become [2]

$$f''' + \frac{3}{4+\gamma} ff'' - \frac{2-\gamma}{4+\gamma} (f')^2 + \tau = 0, \tag{2.11}$$

$$\tau' + \frac{3\sigma\gamma}{4+\gamma} f\tau = 0. \tag{2.12}$$

Boundary conditions in terms of $f$ and $\tau$ are

$$f(0) = f''(0) = \tau'(0) = 0, \tag{2.13}$$

$$f'(\infty) = \tau(\infty) = 0. \tag{2.14}$$

The velocity components are related to $f$ by

$$v_x = a^2 v x^{\frac{2-\gamma}{4+\gamma}} f', \tag{2.15}$$

$$v_y = \frac{-avx^{-\frac{1+\gamma}{4+\gamma}}}{4+\gamma}\left(3f-(1+\gamma)\eta f'\right), \tag{2.16}$$

$$\theta(x,y) = \left(a^4 v^2 \left(g\beta T_\infty^\gamma\right)^{-1}\right)^{1/\gamma} x^{-\frac{3}{4+\gamma}} p(\eta). \tag{2.17}$$

Exact solutions for the system (2.11), (2.12), which satisfy the boundary conditions (2.13) and (2.14), have been found for $\sigma = 2/\gamma$ and $\sigma = (4+\gamma)/9\gamma$. If $\tau$ is assumed related to $f$ by

$$\tau = \frac{5-\gamma}{4+\gamma}(f')^2 + b\left((f')^2 + ff''\right), \tag{2.18}$$

equation (2.11) becomes

$$f''' + \left(b + \frac{3}{4+\gamma}\right)ff'' + \left(b + \frac{3}{4+\gamma}\right)(f')^2 = 0, \tag{2.19}$$

which has the integral

$$f'' + \left(b + \frac{3}{4+\gamma}\right)ff' = 0. \tag{2.20}$$

The same assumption for $\tau$, equation (2.18), is now substituted in the energy equation (2.12), with the result,

$$bf\left(f''' + \frac{3\Pr}{4+\gamma}(ff')'\right) + \frac{10-2\gamma+3b(4+\gamma)}{4+\gamma} f'\left[f'' + \frac{3\gamma(5-\gamma)\Pr}{(4+\gamma)(10-2\gamma+3b(4+\gamma))} ff'\right] = 0. \tag{2.21}$$

Inspection of this equation shows that for certain values of $b$ and $\sigma$ any function which satisfies (2.20) also satisfies (2.21). The appropriate values of $b$ and $\sigma$ are:

$$\text{case (i)} \quad b = 0, \quad \sigma = \frac{2}{\gamma};$$

$$\text{case (ii)} \quad b = \frac{\gamma-5}{3(4+\gamma)}, \quad \sigma = \frac{4+\gamma}{9\gamma}.$$

In case *(i)*, the coefficient of the first bracketed expression in (2.21) vanishes, and the second bracketed expression becomes identical with (2.20), which is, in this

$$f'' + \frac{3}{4+\gamma} ff' = 0 \tag{2.22}$$

The solution of (2.22) and the corresponding expression for $\tau$ from (2.18) are

$$f = \tanh\left(\frac{3}{2(4+\gamma)}\eta\right), \tag{2.23}$$

$$\tau = \frac{9(5-\gamma)}{4(4+\gamma)^3}\sec h^4\left(\frac{3}{2(4+\gamma)}\eta\right). \tag{2.24}$$

In case *(ii)*, the bracketed expressions in (2.21) become identical to the left sides of (2.19) and (2.20). The solution for this case is

$$f = \tanh\left(\frac{1}{6}\eta\right) \tag{2.25}$$

$$\tau = \frac{5-\gamma}{54(4+\gamma)}\sec h^2\left(\frac{1}{6}\eta\right) \tag{2.26}$$

From the solutions for $f$, $f'$ and $p$, one may compute the velocity components and the temperature from equations (2.20), (2.21) and (2.22). However, the constant, $a$, and the location of the origin of $x$ have not as yet been specified. Both of these quantities are related to the initial conditions of the jet, and can be chosen so as to match the mathematical solution to a particular physical experiment.

Suppose that at some station, the velocity and the temperature profiles are measured. Then at some $x$, say $x = x_0$, the mathematical solution must match the measured profiles, at least in an average sense. For instance, if $W$ is the measured volume flow rate, and $E$ is the integral of the measured temperature profile (and is thus a measure of the thermal energy in the jet), one can select $a$ and $x_0$ so as to make the corresponding integrals of the mathematical solution equal to the measured quantities,

$$2\int_0^\infty v_x(x_0, y)\,dy = W, \tag{2.27}$$

$$2\int_0^\infty \left[T(x_0, y) - T_\infty\right]dy = E. \tag{2.28}$$

When equations (2.15) and (2.17) are substituted in (2.27) and (2.28), the results may be solved for $a$ and $x_0$:

$$2ax_0^{\frac{3}{4+\gamma}} = W/\nu, \tag{2.29}$$

$$x_0 = \left(\frac{E}{2}\left(\int_0^\infty \tau^{\frac{1}{\gamma}}(\eta)\,d\eta\right)^{-1}\left(\frac{W}{2}\right)^{\frac{4}{\gamma}}\left(g\beta\nu^2\right)^{-\frac{1}{\gamma}}\right)^{\frac{\gamma(4+\gamma)}{\gamma^2-\gamma-3}}, \tag{2.30}$$

$$a = \frac{W}{2\nu}\left(\frac{E}{2}\left(\int_0^\infty \tau^{\frac{1}{\gamma}}(\eta)\,d\eta\right)\left(\frac{W}{2}\right)^{\frac{4}{\gamma}}\left(g\beta\nu^2\right)^{-\frac{1}{\gamma}}\right)^{\frac{3\gamma}{\gamma^2-\gamma-3}}. \tag{2.31}$$

### 3. The axisymmetric jet

Analysis of the axisymmetric heated jet is completely analogous to the two-dimensional problem. The $z$-axis is now the jet axis, and $r$ is the radial co-ordinate. With the other quantities defined as before, the boundary-layer equations are:

$$\frac{\partial}{\partial z}(rv_z) + \frac{\partial}{\partial r}(rv_r) = 0, \tag{3.1}$$

$$v_z \frac{\partial v_z}{\partial z} + v_r \frac{\partial v_z}{\partial r} = \frac{\nu}{r}\frac{\partial}{\partial r}\left(r\frac{\partial v_z}{\partial r}\right) + g\beta T_\infty^\gamma \theta^\gamma, \tag{3.2}$$

$$v_z \frac{\partial \theta}{\partial z} + v_r \frac{\partial \theta}{\partial r} = \frac{\nu}{\sigma r}\frac{\partial}{\partial r}\left(r\frac{\partial \theta}{\partial r}\right). \tag{3.3}$$

Boundary conditions to be satisfied are:

$$\text{at} \quad r = 0; \quad v_r = \frac{\partial v_z}{\partial r} = \frac{\partial \theta}{\partial r} = 0; \quad v_z \text{ is finite};$$

$$\text{at} \quad r = \infty; \quad v_z = \theta = 0.$$

The continuity equation is again integrated by means of a stream function,

$$v_z = \frac{1}{r}\frac{\partial \psi}{\partial r}, \quad v_r = -\frac{1}{r}\frac{\partial \psi}{\partial z}, \tag{3.4}$$

and introduction of a similarity transformation,

$$\eta = arz^{-\frac{1+\gamma}{4}}, \tag{3.5}$$

$$\psi(z,r) = \nu z f(\eta), \tag{3.6}$$

$$\theta(z,r) = \left(a^4 \nu^2 \left(g\beta T_\infty^\gamma\right)^{-1} \tau(\eta)\right)^{1/\gamma} z^{-1}, \tag{3.7}$$

leads to the ordinary differential equations [2]

$$\left(f'' - \frac{f'}{\eta}\right) + \frac{ff''}{\eta} - \frac{ff'}{\eta^2} - \frac{(1-\gamma)}{2}\frac{(f')^2}{\eta} + \eta\tau = 0, \tag{3.8}$$

$$\eta\tau' + \gamma\sigma f \tau = 0. \tag{3.9}$$

Boundary conditions on $f$ and $\tau$ are:

$$f(0) = f'(0) = \tau'(0) = 0, \tag{3.10}$$

$$\tau(0) \text{ is finite}, \tag{3.11}$$

$$\tau(\infty) = \lim_{\eta \to \infty}\left[f'(\eta)/\eta\right] = 0. \tag{3.12}$$

The velocity components in terms of $f$ are

$$v_z = \frac{a^2 v z^{\frac{1-\gamma}{2}} f'}{\eta} \tag{3.13}$$

$$v_r = \frac{-v}{r}\left(f - \frac{1+\gamma}{4}\eta f'\right) \tag{3.14}$$

Exact solutions for the system (3.8), (3.9), are possible for certain Prandtl numbers. If one assumes

$$f = \frac{b\eta^2}{b+\eta^2}, \tag{3.15}$$

then integration of (3.9) provides $\tau$:

$$\tau = c\left(b+\eta^2\right)^{-\frac{\gamma}{2}\sigma b}. \tag{3.16}$$

Formulas (3.15) and (3.16) satisfy equation (3.8) only for the following choices of $\sigma$, $b$ and $c$:

$$\text{case (i)} \quad \sigma = \frac{3+\gamma}{4\gamma}, \quad b = \frac{24}{3+\gamma}, \quad c = 18432\frac{3-\gamma}{(3+\gamma)^3};$$

$$\text{case (ii)} \quad \sigma = \frac{2}{\gamma}, \quad b = 4, \quad c = 512(3-\gamma).$$

From the solutions for $f$ and $\tau$, one may compute the velocity components and the temperature from equations (3.13), (3.14) and (3.7). The constant $a$, and the location of the origin of $z$; can be chosen in an identical way with that of the two-dimensional jet. Suppose that at some $z$, say $z = z_0$, the velocity and temperature profiles are measured. Then if $W$ is the measured volume flow rate and $E$ is the integral of the measured temperature distribution, one can select $a$ and $z_0$ so as to make the corresponding integrals of the mathematical solution equal to the measured quantities:

$$2\pi \int_0^\infty v_z(z_0, r) r dr = W, \tag{3.17}$$

$$2\pi \int_0^\infty \left[T(x_0, r) - T_\infty\right] r dr = E. \tag{3.18}$$

When equations (3.7) and (3.13) are substituted into (3.17) and (3.18), the resulting equations may be solved for $a$ and $z_0$,

$$z_0 = \left(\frac{W}{2\pi v f(\infty)}\right)^{\frac{2}{1+\gamma}}, \tag{3.19}$$

$$a = (g\beta)^{\frac{2}{2-\gamma}} \left(\frac{E}{2\pi}\right)^{\frac{\gamma}{2(2-\gamma)}} \left(\int_0^\infty \tau(\eta)\eta d\eta\right)^{\frac{\gamma}{2-\gamma}} v^{\frac{1}{2-\gamma}} \left(\frac{W}{2\pi v f(\infty)}\right)^{\frac{\gamma(1-\gamma)}{2(1+\gamma)(2-\gamma)}}. \tag{3.20}$$

The value of $x_0$ now locates the origin of the co-ordinate system, and the mathematical solutions may be assumed to represent the actual jet flow for $x \geq x_0$.

**Acnolegment**